\documentclass[twocolumn,showpacs,pra,superscriptaddress,longbibliography]{revtex4-2}

\usepackage{times}
\usepackage{amsmath}
\usepackage{amssymb}
\usepackage{graphicx}
\usepackage[unicode]{hyperref}
\hypersetup{
   unicode=true,          
   plainpages=false,
   colorlinks=true,       
   citecolor=blue,        
   urlcolor=blue
}
\usepackage[caption=false,position=top,singlelinecheck=off,justification=raggedright]{subfig}
\usepackage{color}
\usepackage{pst-node}

\begin{document}
\title{Dissipative time crystal in an atom-cavity system: Influence of trap and competing interactions}

\author{Richelle Jade L. Tuquero}
\affiliation{National Institute of Physics, University of the Philippines, Diliman, Quezon City 1101, Philippines}

\author{Jim Skulte}
\affiliation{Zentrum f\"ur Optische Quantentechnologien and Institut f\"ur Laser-Physik, 
	Universit\"at Hamburg, 22761 Hamburg, Germany}
\affiliation{The Hamburg Center for Ultrafast Imaging, Luruper Chaussee 149, 22761 Hamburg, Germany}

\author{Ludwig Mathey}
\affiliation{Zentrum f\"ur Optische Quantentechnologien and Institut f\"ur Laser-Physik, 
	Universit\"at Hamburg, 22761 Hamburg, Germany}
\affiliation{The Hamburg Center for Ultrafast Imaging, Luruper Chaussee 149, 22761 Hamburg, Germany}

\author{Jayson G. Cosme}
\affiliation{National Institute of Physics, University of the Philippines, Diliman, Quezon City 1101, Philippines}

\date{\today}
\begin{abstract}

While the recently realized dissipative time crystal in a laser-pumped atom-cavity system in the experiment of Ke{\ss}ler \emph {et al.}~\href{https://journals.aps.org/prl/abstract/10.1103/PhysRevLett.127.043602}{[Phys. Rev. Lett. \textbf{127}, 043602 (2021)]} is qualitatively consistent with a theoretical description in an idealized limit, here, we investigate the stability of this dissipative time crystal in the presence of an inhomogeneous potential provided by a harmonic trap, and competing short- and infinite-range interactions. We note that these features are ubiquitous in any realization of atom-cavity systems. 
By mapping out the dynamical phase diagram and studying how it is modified by the harmonic trap and short-range interactions, we demonstrate the persistence of long-lived dissipative time crystals beyond the idealized limit. We show the emergence of metastable dissipative time crystals with and without prethermalization plateaus for tight harmonic confinement and strong contact interaction, respectively.

\end{abstract}
\maketitle

\section{Introduction}

Time crystals are nonequilibrium many-body phases, in which time-translation symmetry is spontaneously broken \cite{Wilczek2012, Shapere2012, Sacha2020, Else20,Khemani2019}. 
Time crystals that are induced by periodic driving exhibit a robust subharmonic response in relation to the driving frequency. 
Isolated systems under periodic driving, however, continuously heat up, in general, causing the time crystalline order to ``melt" into a featureless state. One approach to stabilize periodically driven time crystals, also known as Floquet time crystals, consists of adding strong disorder to push the system into a many-body localized phase  \cite{Else2016, Yao2017, Khemani2016}. This has enabled the experimental observation of discrete time crystals in various periodically driven systems \cite{Zhang2017,Choi2017,Rovny2018,Randall2021,Mi2022}. Discrete time crystals, which do not rely on many-body localization, have been realized in other experimental platforms \cite{Smits2018,Autti2018,Monroe2021}.
Alternative strategies to stabilize time crystals include coupling the system to an environment \cite{Else2017,Gong2018,Iemini2018,Buca2019,Michal2022} or including long-range interaction \cite{Russomanno2017,Kozin2019,Kelly2021,Pizzi2021a,Ye2021}. 

Dissipation, in particular, has been utilized to create a dissipative time crystal (DTC) in an atom-cavity system \cite{Kessler2021}.  Due to the approximation of the atom-cavity system via the Dicke model \cite{Mivehvar2021}, this paradigmatic DTC can be regarded as a realisation of the Dicke time crystal \cite{Gong2018}. The open Dicke model describes an ensemble of two-level systems interacting with photons in a leaky cavity \cite{Kirton2019}. Note that the standard Dicke model does not have any notion of spatial dimension as it is a zero-dimensional model. That is, it excludes spatially dependent potential for the atoms. More importantly, it only captures the all-to-all photon-mediated coupling between the atoms. 
These approximations provide an idealized limit, in which the spatial $\mathbb{Z}_2$ symmetry breaking in the superradiant phase is intimately tied to the temporal $\mathbb{Z}_2$ symmetry breaking of the period-doubled Dicke time crystal. Moreover, the infinite-range nature of the cavity-mediated interaction makes this approximate description mean-field solvable and, in fact, exactly solvable in the thermodynamic limit \cite{Gong2018,Zhu2019}. 
\begin{figure}[!htb]
\centering
\includegraphics[width=1\columnwidth]{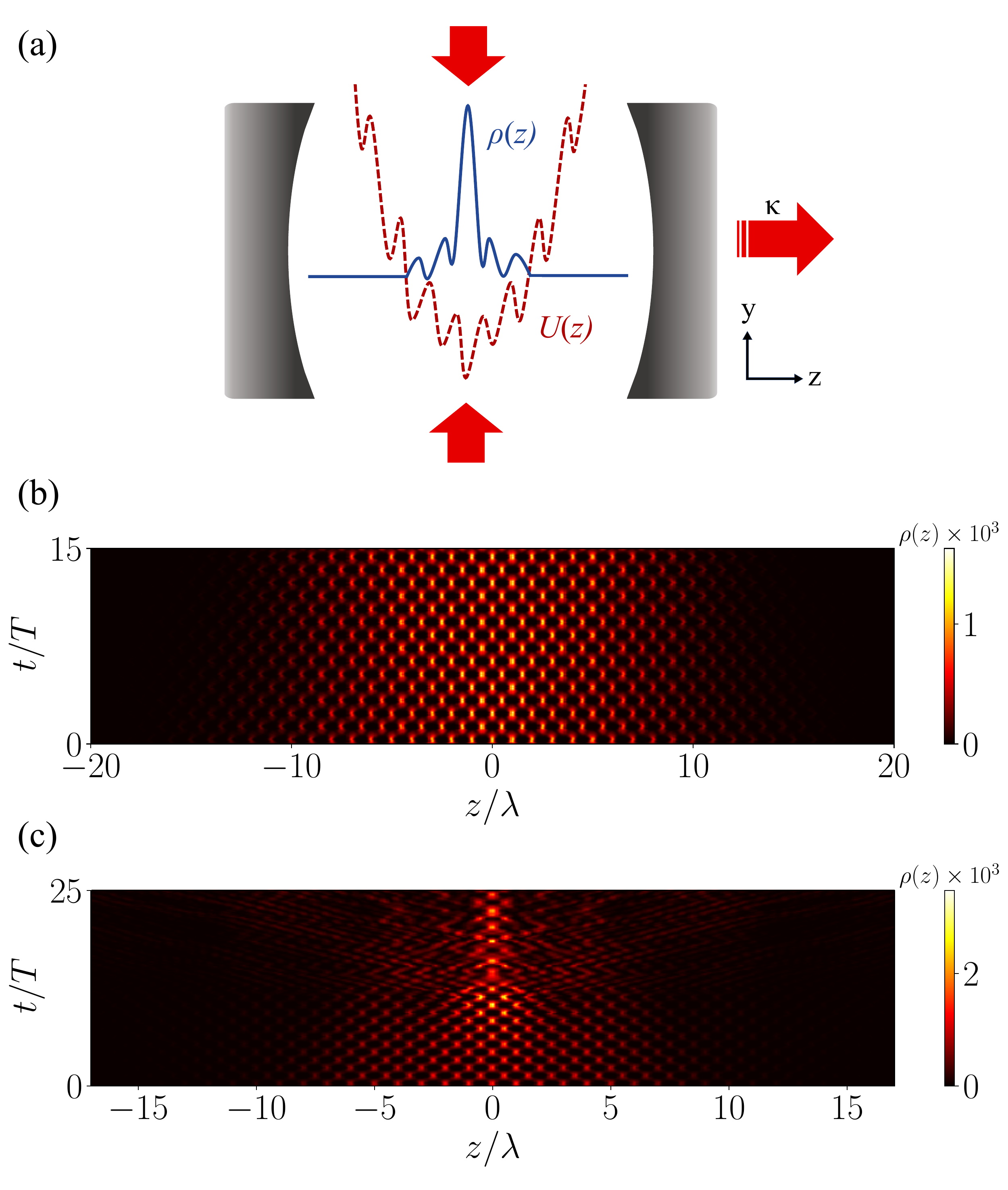}
\caption{(a) Schematic diagram of the atom-cavity system consisting of a Bose-Einstein condensate with a harmonic trap inside a high-finesse optical resonator. The solid curve denotes the particle distribution $\rho(z)$ in the self-organized density wave phase and the dashed curve denotes the combined dipole potential $U(z)$ due to the cavity field and the harmonic trap. The cavity photon loss rate is $\kappa$. The pump and cavity wavelength is $\lambda$. Dynamics of the atomic density for a (b) stable and (c) metastable dissipative time crystal with harmonic trap frequency $\omega = \hbar/m(3.5\lambda)^2$ and short-range interaction energy $E_\mathrm{int}/E_\mathrm{rec} = 0.26$ with $E_\mathrm{int}$ as defined Sec.~\ref{sec:dia}.  In (b), the driving strength is $f_\mathrm{d} = 0.7$ and the driving frequency is $\omega_\mathrm{d}/2\pi = 5$ kHz. While in (c), $f_\mathrm{d} = 0.5$ and $\omega_\mathrm{d}/2\pi = 3.5$ kHz.}
\label{fig:schem} 
\end{figure} 
The absence of an inhomogeneous potential and other forms of interaction distinguishes the Dicke model from an atom-cavity system in Ref.~\cite{Kessler2021} and any other realisations, where an external harmonic trap and inherent collisional interactions between the atoms are present. These subtle yet important distinctions pose the question about the stability of DTCs, or any prediction based on the Dicke approximation, in realistic atom-cavity setups, in addition to the finite lifetime of Bose-Einstein condensates. 
On the one hand, inhomogeneous potentials, such as a harmonic trap, break the spatial symmetry, which could affect the $\mathbb{Z}_2$ symmetry broken states responsible for the period-doubling response. On the other hand, short-range interactions break the mean-field solvability of the Dicke model \cite{Zhu2019}.  Given that these features are ubiquitous in any atom-cavity system, it is imperative to point out their influence. 
In the absence of dissipation, beyond mean-field effects on discrete time crystals in a spin model with competing short- and long-range interactions are studied in Ref.~\cite{Pizzi2021b}.

In this work, we investigate the influence of harmonic confinement and short-range interaction on the DTC found in a periodically driven atom-cavity system. We show that a DTC remains stable in the presence of weak perturbations that explicitly break spatial symmetry and mean-field solvability of the model. 
This is in contrast to the absence of a stable period-doubling response predicted in a similar atom-cavity setup but for the bad cavity limit, wherein the cavity dynamics is orders of magnitude faster than the atomic dynamics  \cite{Molignini2018}. 
While it was mentioned in Ref.~\cite{Kessler2021} that strong contact interaction may lead to a DTC with finite lifetime, similar to a metastable Dicke time crystal \cite{Zhu2019}, a detailed analysis of this phase, which we call metastable DTC, is still lacking. 
Aside from the Dicke time crystal, other dynamical phases in dissipative systems are found to exhibit metastability when pushed out of the idealized limit \cite{Sarkar2021,Jamir2022}. 
Here, we show that metastable DTCs may emerge not only because of short-range interaction competing with the infinite-range interaction but also due to the influence of harmonic confinement. 

This work is organized as follows. We discuss the system, the method for simulating the dynamics, and the driving protocol in Sec.~\ref{sec:system}. The properties of the stable DTC in the ideal atom-cavity system is reviewed in Sec.~\ref{sec:ideal}. In Sec.~\ref{sec:dia}, we map out and analyze the dynamical phase diagram for different combinations of contact interaction strength and harmonic trap frequency. In Sec.~\ref{sec:mdtc}, we further study the metastable dissipative time crystal and its lifetime. Finally, we conclude this paper in Sec.~\ref{sec:conc}

\section{System}\label{sec:system}

We consider an atom-cavity system with a Bose-Einstein condensate (BEC) of $^{87}$RB atoms as depicted in Fig.~\ref{fig:schem}(a). An external laser is applied along the $y$ direction, which is perpendicular to the cavity axis aligned in the $z$ direction. Photons leak out of the cavity at a rate of $\kappa$. The cavity wavelength is $\lambda$. An external harmonic trap is present along the $z$ direction. In the following, we consider the one-dimensional limit of the system and investigate only the dynamics along the cavity axis.  

We vary the pump intensity $\epsilon$ to investigate the dynamical response of the system. The transversely pumped atom-cavity system hosts a self-organisation phase transition \cite{Ritsch2013}. Above a critical value $\epsilon_\mathrm{crit}$, it becomes energetically favourable for the atoms to self-organise into a chequerboard density wave (DW) phase to scatter photons from the pump into the cavity \cite{Baumann2010, Klinder2015}. This self-organisation phase transition is as an approximate emulation of the superradiant phase transition in the Dicke model \cite{Kirton2019}.
In the density wave phase, the system breaks the $\mathbb{Z}_2$ symmetry as the atoms spontaneously localise either in the odd or even sites of the emergent standing wave formed by the cavity photons. These two symmetry broken states can be distinguished by the sign of the density wave order parameter $\Theta= \langle \mathrm{cos}(kz)\rangle$ where $k=2\pi/\lambda$. That is, a non-zero positive (negative) value for $\Theta$ corresponds to an even (odd) DW state \cite{Ritsch2013, Nagy2008, Baumann2010, Klinder2015}. 

The Hamiltonian for the system is a combination of the cavity and the atom Hamiltonian, as well as the short-range atom-atom interaction and the atom-cavity interaction, i.e.,
\begin{equation}\label{H}
	\hat{H} = \hat{H}_\mathrm{C}+\hat{H}_\mathrm{A}+\hat{H}_\mathrm{AA}+\hat{H}_\mathrm{AC}.
\end{equation}
The Hamiltonian for the cavity with a mode function $\cos(kz)$ is given by
\begin{equation}\label{Hc}
	\hat{H}_\mathrm{C} = -\hbar\delta_\mathrm{C}\hat{\alpha}^\dagger \hat{\alpha}. 
\end{equation}
where $\delta_\mathrm{C}$ is the pump-cavity detuning and $\hat{\alpha}^\dagger$ ($\hat{\alpha}$) is the creation (annihilation) operator for the cavity photon.
We include a harmonic trap for the atoms with trap frequency $\omega$. The single-particle Hamiltonian for the atoms is
\begin{equation}\label{Ha}
	\hat{H}_\mathrm{A} = \int \hat{\Psi}^\dagger(z) \left[ -\frac{\hbar^2}{2m}\frac{d^2}{dz^2}+\frac{1}{2}m\omega^2 z^2\right]\hat{\Psi}(z) \;dz
\end{equation}
where $m$ is the mass of a $^{87}$Rb atom and $\Psi(z)$ is the bosonic field operator associated with the BEC.
We are interested in the interplay between the infinite-range cavity-mediated interaction and the inherent short-range collisional interaction between the atoms. The short-range interaction is described by
\begin{equation}
	\hat{H}_\mathrm{AA} = \frac{g_\mathrm{aa}}{2} \int \hat{\Psi}^\dagger (z) \hat{\Psi}^\dagger(z) \hat{\Psi} (z) \hat{\Psi}(z)\;dz.\label{Haa}
\end{equation}
where $g_\mathrm{aa}$ is the contact interaction strength. On the other hand, the atom-cavity interaction, which gives rise to a dynamical infinite-range interaction between the atoms, is modeled by
\begin{align}
		\hat{H}_\mathrm{AC} =\int \hat{\Psi}^\dagger(z) \hbar U_0  [\mathrm{cos}^2(kz)\hat{\alpha}^\dagger \hat{\alpha} \label{Hac}\\ \nonumber
		 +\sqrt{\frac{{\epsilon}}{\hbar|U_0|}} \mathrm{cos}(kz)(\hat{\alpha}^\dagger +\hat{\alpha})]\hat{\Psi}(z)\; dz.
\end{align}
The pump frequency is red-detuned with respect to the atomic transition frequency leading to a negative light shift per photon $U_0<0$. Note that the atom-cavity interaction strength depends on $U_0$ and the pump intensity $\epsilon$.

The dynamics of the system is captured by the following Heisenberg-Langevin equations
\begin{align}\label{eq:eom}
	\frac{\partial}{\partial t}\hat{\Psi} &= \frac{i}{\hbar}[\hat{H}, \hat{\Psi}] \\ 
	\frac{\partial}{\partial t}\hat{\alpha} &= \frac{i}{\hbar}[\hat{H}, \hat{\alpha}]-\kappa \hat{\alpha}+\xi,
\end{align}
where $\xi$ is the stochastic noise due to the cavity dissipation with $\langle \xi^*(t)\xi(t')\rangle = \kappa \delta(t-t')$ \cite{Ritsch2013}.
We employ the truncated Wigner approximation (TWA) to simulate the quantum dynamics \cite{Polkovnikov2010, Carusotto2013}. The TWA goes beyond the mean-field approximation through the inclusion of quantum noise from the initial state and the fluctuations corresponding to the dissipation in the cavity. This method treats the quantum operators as $c$ numbers and it is applicable for large number of atoms and weak coupling. For the initial states, we choose coherent states for the BEC and the empty cavity mode. 
We then propagate an ensemble of initial states, which samples the initial Wigner distributions, according to the coupled stochastic differential equations in Eq.~\eqref{eq:eom}. The TWA has been used to confirm robustness of dissipative time crystals \cite{Cosme2019,Kessler2019,Kessler2020} and in comparison with experiment \cite{Kessler2021,Popla2022}.

We assume the initial state of the BEC to be homogeneous in the absence of a harmonic trap. When a harmonic trap is present, we use imaginary time propagation to initialize the system in the ground state (see Appendix \ref{sec:imag} for details). We consider the pump protocol depicted in Fig.~\ref{fig:cleanDTC}(a). The pump intensity is linearly increased for 2.5 ms until it reaches $\epsilon_0 = 1.02 \epsilon_\mathrm{crit}$, where $\epsilon_\mathrm{crit}$ is the critical pump intensity for self-organisation for a given contact interaction strength and harmonic oscillator frequency. Next, $\epsilon$ is held constant until 30 ms allowing the system to relax to the corresponding DW state. Finally, the pump intensity is periodically modulated according to
\begin{equation} 
	\epsilon (t) = \epsilon_0(1 + f_\mathrm{d} \sin(\omega_\mathrm{d}t)),
\end{equation}
where $f_\mathrm{d}$ is the driving strength and $\omega_\mathrm{d}$ is the driving frequency. The driving period is $T=2\pi/\omega_{\mathrm{d}}$.

In the following, we use realistic parameters according to the experimental set-up in Ref.~\cite{Kessler2021}. The particle number is $N_\mathrm{a} = 65 \times 10^3$ atoms,  recoil frequency $\omega_\mathrm{rec} = 2\pi^2\hbar/m\lambda^2 = 2\pi \times 3.55$ kHz, decay rate $\kappa = 2\pi \times 4.55$ kHz, $U_0 = -2\pi \times 0.36$ Hz, and effective pump detuning $\delta_\mathrm{eff} = \delta_\mathrm{C}-N_\mathrm{a}U_0/2 =-2\pi \times 18.5$ kHz. We simulate the dynamics for 200 driving cycles.

\section{Ideal Dissipative Time Crystal}\label{sec:ideal}

In this section, we recall the defining properties of a DTC in the ideal limit when both contact interaction and harmonic trap are absent, as a preparational step to determine their influence in Sec.~\ref{sec:dia}. In this limit, the atom-cavity system maps approximately onto the Dicke model if inhomogeneous trap and contact interactions are neglected, and thus, the DTC observed here is equivalent to the paradigmatic Dicke time crystal \cite{Gong2018}. 
\begin{figure}[!htb]
	\centering
	\includegraphics[width=1\columnwidth]{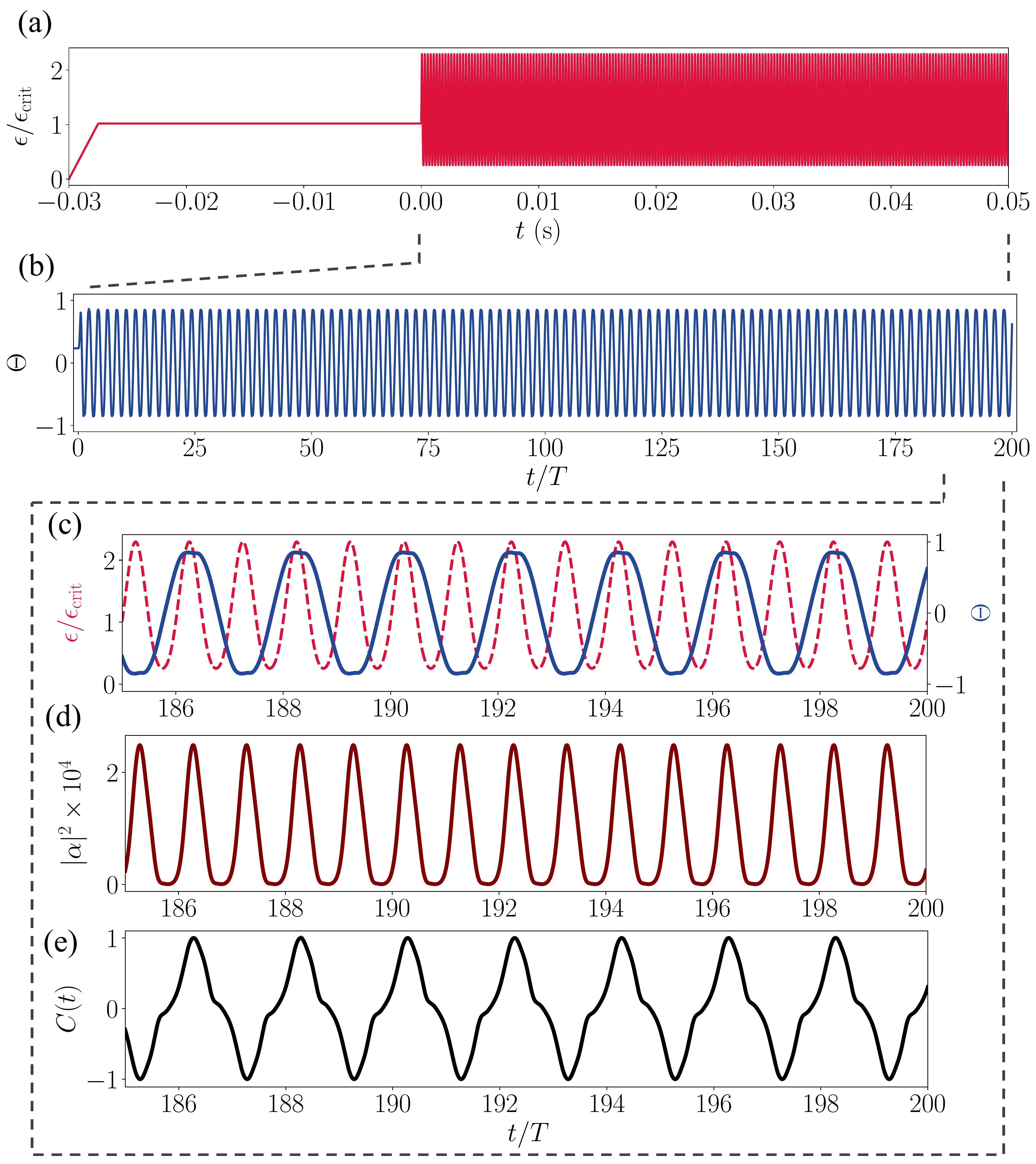}
	\caption{(a) Protocol for the pump intensity. Dynamics of a single TWA trajectory for the (b) order parameter during periodic modulation. Zoom-in of the dynamics for the (c) order parameter, (d) intracavity photon number $|\alpha|^2$, and (e) the correlation function $C(t)$. The driving parameters are $f_\mathrm{d} = 0.5$ and $\omega_\mathrm{d}/2\pi = 4$ kHz in the absence of a harmonic trap and contact interactions.}
	\label{fig:cleanDTC} 
\end{figure} 

Modulation of the pump intensity leads to the formation of a DTC in the atom-cavity system, for a specific regime of driving strength and frequency \cite{Kessler2021,Cosme2019}. This dynamical phase is characterized by a period-doubled switching between the symmetry broken DW states. The mean-field approximation of the ideal DTC is depicted in Fig.~\ref{fig:cleanDTC}(b-c). The periodic switching of the sign of the order parameter in Fig.~\ref{fig:cleanDTC}(c) underpins how the system switches between the odd and even DW states. Moreover, the switching occurs at twice the driving period as seen in Fig.~\ref{fig:cleanDTC}(c). Another important characteristic of a DTC is seen from the dynamics of the cavity mode occupation $|\alpha|^2$, which exhibits pulsating behavior at the driving frequency, see Fig.~\ref{fig:cleanDTC}(d). This means that the DTC rely on the presence of cavity photons, which mediate an infinite-range interaction between the atoms, and thus highlights the many-body nature of the DTC phase.

To quantify the behavior of the time crystal using the TWA, we obtain the two-point temporal correlation function 
\begin{equation}
C(t) = \mathrm{Re}\{\langle \hat{\alpha}^\dagger(t)  \hat{\alpha}(t_0)\rangle\}/\langle  \hat{\alpha}^\dagger(t_0)  \hat{\alpha}(t_0)\rangle,
\end{equation}
where $t_0$ is the time before modulation is switched on. In Fig.~\ref{fig:cleanDTC}(e), we present an example of the dynamics of $C(t)$ in a ideal DTC. Note that it closely follows the behavior of the order parameter $\Theta$. In the following, we use $C(t)$ instead of $\Theta$, which averages out in TWA due to the $\mathbb{Z}_2$ symmetry breaking response of the DTC.

\section{Dynamical Phase Diagrams}\label{sec:dia}
\begin{figure*}[!htp]
	\centering
	\includegraphics[width=2\columnwidth]{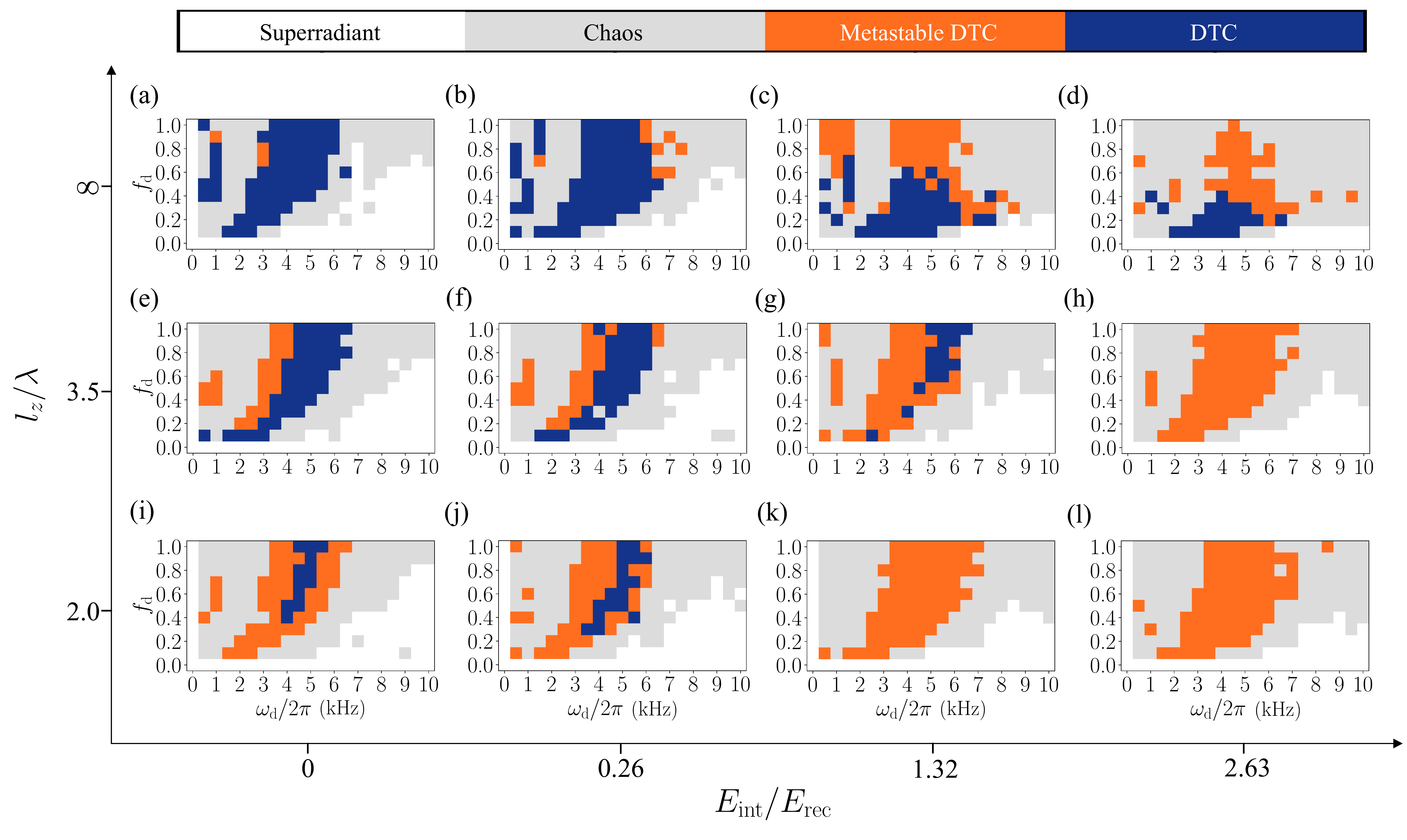}
	\caption{Gallery of dynamical phase diagrams for increasing contact interaction strength, from left (zero contact interaction) to right, and increasing harmonic oscillator trap frequency, from top (no harmonic trap, $\omega=0$) to bottom. The oscillator strengths in units of recoil energy from top to bottom are $E_\mathrm{osc}/E_\mathrm{rec}=\{0,0.004,0.013\}$.}
	\label{fig:phases} 
\end{figure*} 
We now explore the influence of harmonic confinement and short-range interactions between the atoms on the stability of the DTC. 
The harmonic trap frequency $\omega$ is related to the oscillator length $l_z$ via $\omega = \hbar/(l_z^2 m)$. Alternatively, we measure the confinement strength by comparing $E_\mathrm{osc} = \hbar \omega $ with the recoil energy $E_\mathrm{rec} = \hbar \omega_\mathrm{rec}$
We quantify the contact interaction via the mean-field interaction energy $E_\mathrm{int} = g_\mathrm{aa} N_\mathrm{a}/\lambda$, where $g_\mathrm{aa}>0$ is the repulsive contact interaction strength. 

We use $C(t)$ to classify the phases in the dynamical phase diagrams shown in Fig.~\ref{fig:phases}. Specifically, we classify a stable or persistent DTC if $C(t)$ perfectly switches sign every driving cycle for the final 100 driving periods. 
We also observe the emergence of a metastable DTC phase. On the level of a single TWA trajectory, we define a metastable DTC by having a $C(t)$ that switches sign at least six consecutive times (or equivalently three consecutive period doubling) during the initial stage of driving, $t\in[0,6T]$, before eventually becoming chaotic, in which $C(t)$ does not change sign over multiple driving periods with irregular intervals. We identify a completely chaotic phase by the lack of consecutive period doubling over $t\in[0,6T]$ in addition to the obvious irregular dynamics of $C(t)$  (see Appendix \ref{sec:chaos} for an example). In Fig.~\ref{fig:phases}, the dynamical phase diagrams are arranged in increasing contact interaction strength from left to right and increasing harmonic oscillator frequency from top to bottom. The resonant nature of DTCs in atom-cavity systems \cite{Kessler2021,Skulte2021,Popla2021} is evident from the fact that both stable and metastable DTCs are only found in some range of the driving frequency in Fig.~\ref{fig:phases}. 

The dynamical phase diagram in the ideal scenario, in which both harmonic trap and contact interaction are neglected is shown in Fig.~\ref{fig:phases}(a). A stable DTC can be found in a large area of the driving parameter space, specifically for $\omega_d/2\pi \in [1,6]~\mathrm{kHz}$. We also observe a DTC in a much smaller region of the parameter space for low driving frequencies, $\omega_d/2\pi < 1.0~\mathrm{kHz}$. This phase is distinct from the usual DTC phase due to the presence of faster but subdominant oscillations in the order parameter corresponding to third harmonics as exemplified in Appendix \ref{sec:dtclow}. While this phase is robust against the quantum noise included in TWA, it is noticeably less robust against contact interaction and harmonic confinement as inferred from the gradual disappearance of the relevant region in Fig.~\ref{fig:phases}.

\begin{figure*}[!htbp]
	\centering
	\includegraphics[width=2\columnwidth]{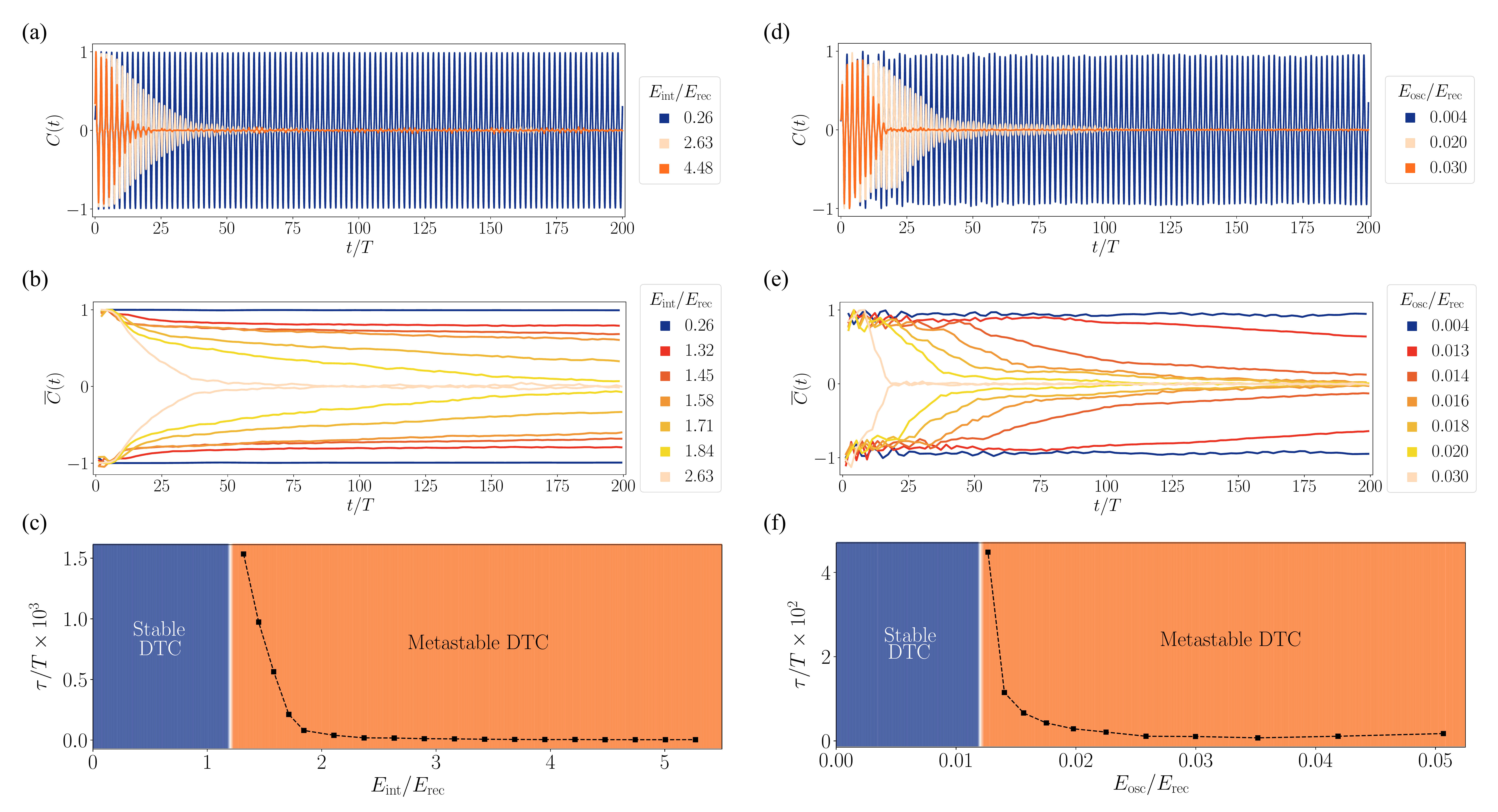}
	\caption{(a)-(c) Behaviour of the correlation function for varying contact interaction strength $(E_\mathrm{int}>0)$ and zero oscillator frequency $(E_\mathrm{osc}=0)$ obtained using TWA with $10^3$ trajectories. (a) Exemplary dynamics of $C(t)$. (b) Stroboscopic correlation function $\overline{C}(t)$ for different contact interaction strength. (c) Dependence of the lifetime $\tau$ on the harmonic oscillator frequency. The driving parameters are $f_\mathrm{d} = 0.5$ and $\omega_\mathrm{d}/2\pi = 4$ kHz. (d)-(f) Similar to (a)-(c) but for varying harmonic trap frequency $(E_\mathrm{osc}>0)$ and zero contact interaction $(E_\mathrm{int} = 0)$.}
	\label{fig:lz} 
\end{figure*} 

We find stable DTC in the presence of inhomogeneous trapping and short-range interaction. This suggests that indeed a stable DTC phase can form in the atom-cavity system with harmonic trap and inherent collisional interaction, and thus supports the experimental observation of a DTC in Ref.~\cite{Kessler2021}. 
Moreover, the presence of both short-range collisional interaction and infinite-range cavity-mediated interaction between the atoms implies the departure of the DTC from the mean-field regime.
This persistence of the DTC phase agrees with the prediction of a long-lived Dicke time crystal despite the presence of short-range interaction, which breaks the mean-field solvability of the Dicke model \cite{Zhu2019}. Typical dynamics of the atomic distribution in the  stable DTC phase in the nonideal limit is demonstrated in Fig.~\ref{fig:schem}(b), which shows the system periodically switching between the odd and even DW states.

The parameter regime with a stable DTC phase decreases with increasing short-range interaction strength and harmonic confinement as seen in Fig.~\ref{fig:phases}. The stable DTC is replaced by either a chaotic phase or a metastable DTC.  In general, the region of the chaotic phase expands with increasing $E_\mathrm{int}$, which is a consequence of the nonlinear nature of the short-range interaction that couples the periodic motion to a continuum of excitations of the atomic cloud. Strong driving enables a DTC with large photon number and deep intracavity field, thereby forming large density modulations in the atomic distribution (see Appendix \ref{sec:dtcfd}). 
The energy associated with the collisional interaction is large for a distribution with large density modulation, which means that repulsive contact interaction penalizes its formation. As demonstrated in Figs.~\ref{fig:phases}(a)-(d), this leads to the suppression of stable DTCs in the strong driving regime, $f_\mathrm{d} > 0.5$, for increasingly strong contact interaction.

In addition to contact interaction, strong harmonic confinement can also destabilize a DTC due to trap-induced coupling between relevant momentum modes becoming dominant over cavity-induced coupling, see Appendix \ref{sec:coup}. 
Note that strong harmonic confinement increases the density at the center of the trap, while strong contact interaction reduces it. These two system properties therefore have competing influence on the density.
This explains how strong harmonic confinement ``melts" the DTCs with small density modulation corresponding to weak driving strength $f_\mathrm{d}$, as demonstrated in Fig.~\ref{fig:phases}(i), which is in contrast to the effect of contact interaction shown in Fig.~\ref{fig:phases}(c). Because of their competing effect on the particle density, one may have naively expected that the contact interaction may be tuned appropriately to cancel the effect of the harmonic trap and therefore stabilize the DTC phase. 
However, we highlight in Figs.~\ref{fig:phases}(e)-(h) and \ref{fig:phases}(i)-(l) that this is not the case and, in fact, increasing the contact interaction strength leads to further destabilisation of the DTC. 
Similarly, for a fixed contact interaction strength, tightening the trap in an attempt to counteract the repulsive interaction shrinks the area in the phase diagram where a stable DTC persists, as seen in Figs.~\ref{fig:phases}(b), \ref{fig:phases}(f), and \ref{fig:phases}(j). The harmonic trap or any inhomogeneous potentials, in general, will couple various momentum modes. Such a coupling may significantly deplete the momentum modes that are important for the $\mathbb{Z}_2$ symmetry broken states participating in the DTC phase, namely the $|\mathbf{k}=0\rangle$ and $|\mathbf{k}=2\pi/\lambda\rangle$ momentum modes. Thus, we demonstrate the importance of ensuring a weak inhomogeneous trap to obtain a stable DTC.

\section{Metastable Dissipative Time Crystal}\label{sec:mdtc}

We now further investigate the metastable dissipative time crystal, the predominant nontrivial dynamical phase for large oscillator frequency and large contact interaction strength, as seen in the phase diagrams in Fig.~\ref{fig:phases}. The metastable DTC exhibits a period-doubled response on a time scale larger than the oscillation period before the dynamics become irregular as shown in Fig.~\ref{fig:schem}(c), for example.

We first focus on the case without a harmonic trap but with a nonzero short-range interaction. In the metastable DTC phase, the irregularity in the dynamics for a single trajectory translates into an exponentially decaying oscillations of the temporal correlation $C(t)$ after averaging over multiple trajectories in TWA as shown in Fig.~\ref{fig:lz}(a). Moreover, we obtain the stroboscopic correlation function $\overline{C}(t)$ defined as the envelope of the oscillations in the correlation function. 
A metastable DTC is characterized by having a finite lifetime, $\tau$, which we extract by fitting an exponential decay $\sim \exp(-t/\tau)$ to the corresponding stroboscopic correlation function $\overline{C}(t)$, as depicted in Fig.~\ref{fig:lz}(b). 
We demonstrate in Fig.~\ref{fig:lz}(c) that the lifetime decreases with the contact interaction strength. Similar to the metastable Dicke time crystal \cite{Zhu2019}, we emphasise that the metastable DTC for $E_\mathrm{int}>0$ without harmonic confinement is distinct from the prethermal discrete time crystals, which rely on high driving frequency to increase the relaxation time towards a featureless thermal state \cite{Machado2019,Pizzi2021,Monroe2021}. 
Unlike in a prethermal discrete time crystal,  there is no visible prethermalization plateaus in Fig.~\ref{fig:lz}(b) for the metastable DTCs induced by contact interaction.

Next, we present in Fig.~\ref{fig:lz}(d) the representative dynamics of $C(t)$ when there is a harmonic confinement but short-range interaction is ignored. The fluctuation in the oscillation amplitude of $C(t)$ is more pronounced during the initial dynamics, $t/T\in[0,20]$, but it stabilizes in the long-time limit for a stable DTC under weak confinement. The fluctuating oscillation amplitude is highlighted in the relatively noisy dynamics of the stroboscopic correlation functions shown in Fig.~\ref{fig:lz}(e). The oscillations inferred from Fig.~\ref{fig:lz}(b) are more stable compared to those in Fig.~\ref{fig:lz}(e), which corroborates the role of the harmonic trap in introducing small irregularity in the period-doubling response and the photon number dynamics observed in the experiment \cite{Kessler2021}.
In Fig.~\ref{fig:lz}(f), we find that the lifetime of a metastable DTC decreases with increasing harmonic oscillator frequency similar to the effect of the short-range interaction. 
Both energy scales Figs.~\ref{fig:lz}(c) and \ref{fig:lz}(f) are in relation to the recoil energy, i.e. $E_\mathrm{osc}/E_\mathrm{rec}$, and $E_\mathrm{int}/E_\mathrm{rec}$. We further observe that the typical lifetime of metastable DTCs with harmonic confinement is shorter by an order of magnitude than those without the trap but with contact interaction, as inferred from comparing Figs.~\ref{fig:lz}(c) and \ref{fig:lz}(f). 
This implies that an inhomogeneous potential, such as the harmonic trap, has a more detrimental effect on the stability of dissipative time crystals that rely on states with spatial long-range order, like the DW phase in the atom-cavity system. 

A different kind of metastable DTC emerges for strong harmonic confinement without contact interaction. We observe in Fig.~\ref{fig:lz}(e) the appearance of prethermalization plateaus, wherein the amplitude of the period-doubling response is fixed, reminiscent of those found in prethermal discrete time crystals \cite{Machado2019,Pizzi2021,Monroe2021}. This behavior is different from the exponential decay observed as soon as the periodic driving starts in the \textit{standard} metastable DTCs for strong contact interactions and without harmonic trap, see Fig.~\ref{fig:lz}(b). Thus, we propose a second kind of metastable DTC, a prethermal dissipative time crystal (PDTC). This phase can be understood in the paradigm of prethermalization arising from fast driving. In the presence of a harmonic trap, the energy of the system can be rescaled by the oscillator frequency $\omega$. Then, the ratio between the driving frequency and the oscillator frequency becomes the relevant energy scale for defining the ``fast-driving regime", $\omega_\mathrm{d}/\omega \gg 1$. Similar to a prethermal discrete time crystal \cite{Machado2019,Pizzi2021,Monroe2021}, the relaxation time in the PDTC can be increased by increasing the relative driving frequency, $\omega_\mathrm{d}/\omega$, which is effectively achieved by decreasing the oscillator frequency as demonstrated in Fig.~\ref{fig:lz}(e). This point of view is consistent with the infinitely long-lived DTC found in the mean-field limit $\omega = 0$, in which $\omega_\mathrm{d}/\omega \to \infty$.


\section{Conclusions}\label{sec:conc}

In conclusion, we have investigated the properties of dissipative time crystals under realistic conditions of the atom-cavity system. Specifically, we included a harmonic confining potential and short-range interactions that compete with the infinite-range interaction mediated by the cavity photons. 
Our results demonstrate that the DTC phase is robust for a nonzero harmonic potential and contact interaction strength, consistent with the observation of a DTC in a similar setup \cite{Kessler2021}. 
We also show that the irregular amplitude of oscillations observed in the experiment \cite{Kessler2021} can be attributed to the harmonic trap.
We point out that for the bad cavity $\kappa \gg \omega_\mathrm{rec}$, and similar conditions, there seems to be no evidence for a stable DTC phase \cite{Molignini2018}, which may hint at the importance of having recoil resolution $\kappa \sim \omega_\mathrm{rec}$ as considered in this work and in the experimental setup in Ref.~\cite{Kessler2021}. 

For sufficiently strong harmonic confinement and contact interaction, a DTC may become unstable towards the formation of two kinds of metastable DTC. Strong contact interactions lead to an exponentially decaying period-doubling response. On the other hand, strong harmonic trap gives rise to prethermal DTC with prethermalization plateaus, during which the correlation function exhibits subharmonic response at a fixed amplitude.
Our work sheds light on the crucial role of trapping on the stability of a DTC, which we expect to apply to inhomogeneous potentials, in general. 
The metastable DTCs are a genuine many-body phase produced by the interplay between driving, dissipation, and mean-field breaking effects, such as competing range of interaction and inhomogeneity in space. We provide not only a strategy for stabilising DTC but also a route for systematic realization of a standard metastable DTC and a prethermal DTC. As an outlook, the transition from a stable to a metastable DTC can be experimentally explored using a combination of Feshbach resonance for tuning the contact interaction strength \cite{Chin2010} and digital micromirror device for creating arbitrary potential for the atoms \cite{Gauthier16}. Finally, we emphasise all atom-cavity systems have a confining potential and atomic interactions.  Our study demonstrates a general strategy to determine the influence of these inevitable features on any many-body state that is created in these systems.

\begin{acknowledgments}
R.J.L.T. and J.G.C. acknowledge support from the DOST-ASTI's COARE high-performance computing facility. J.S. acknowledges support from the German Academic Scholarship Foundation. L.M. and J.S. are supported  by the Deutsche Forschungsgemeinschaft (DFG) in the framework of the Cluster of Excellence “Advanced Imaging of Matter” (EXC 2056), Project No. 390715994. L.M. is also supported by the DFG in the framework of SFB 925, Project No. 170620586.
\end{acknowledgments}

\setcounter{equation}{0}
\setcounter{table}{0}
\appendix
\section{Initial ground state}\label{sec:imag}

We use imaginary time propagation $t \rightarrow -it$ in the underlying equations of motion to initialize the system in the ground state with harmonic trap and contact interaction. To check if our scheme works, we also propagate the system in real time with the same harmonic trap and contact interaction, which in principle should render the density profile unchanged. 
\begin{figure}[!htb]
	\centering
	\includegraphics[width=1\columnwidth]{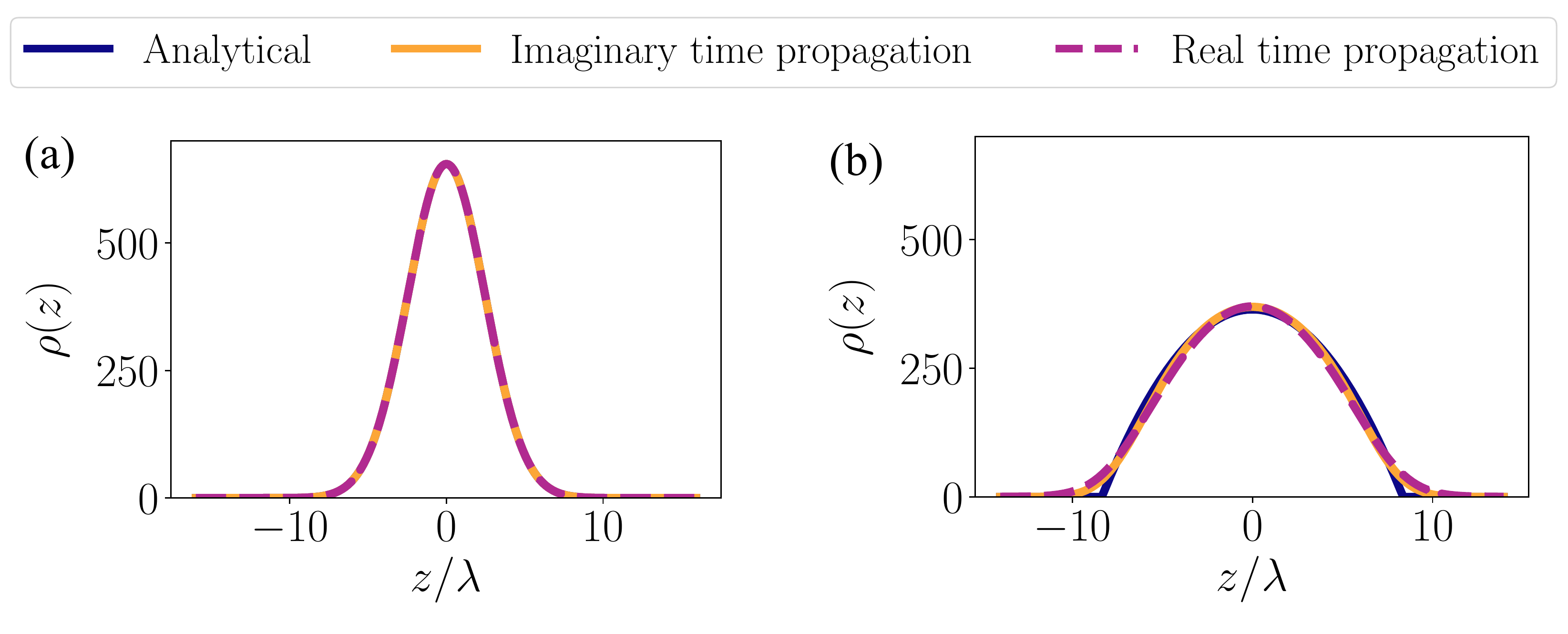}
	\caption{Density profile according to the analytical prediction, imaginary time propagation, and real time propagation for $l_z/\lambda = 3.5$  with (a) zero short-range contact interaction and (b) $E_\mathrm{int}/E_\mathrm{rec} = 0.2634$. In (a), the analytical density profile is Gaussian function, while in (b) it follows from Thomas-Fermi approximation.}
	\label{fig:ground} 
\end{figure} 
As an initial guess for the case when there is only a harmonic trap, we use the exact ground state of the quantum harmonic oscillator, which is a Gaussian function. Fig.~\ref{fig:ground}(a) confirms the validity of the imaginary time propagation as the exact analytical ground state matches both the density profiles obtained from the imaginary and real time propagation methods. On the other hand, for nonzero contact interaction, the appropriate initial guess is the density profile according to the Thomas-Fermi approximation \cite{Pethick}. Fig.~\ref{fig:ground}(b) displays good agreement between the three methods.

\section{Chaotic phase}\label{sec:chaos}
\begin{figure}[!htb]
	\centering
	\includegraphics[width=0.9\columnwidth]{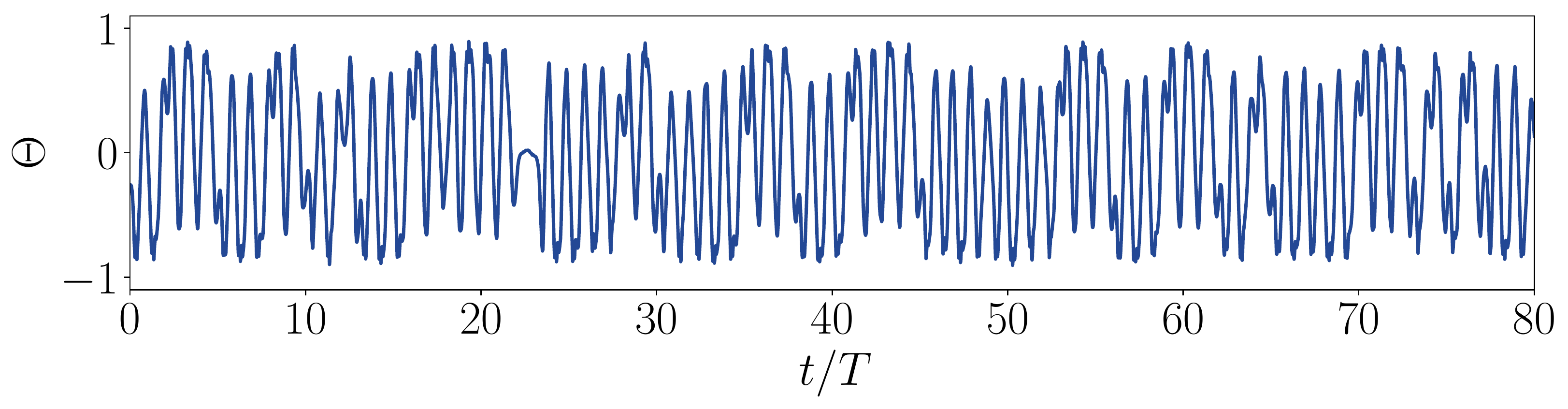}
	\caption{Time evolution of the order parameter $\Theta$ in the chaotic phase. The driving parameters are $f_\mathrm{d}=0.8$ and $\omega_\mathrm{d}/2\pi =2.5$ kHz.}
	\label{fig:chaos} 
\end{figure} 
The chaotic phase is characterized by intermittent dynamics as the system gets stuck in one of the DW states randomly in time. In Fig.~\ref{fig:chaos}, we show the dynamics of the order parameter in the chaotic phase.

\section{DTC for low driving frequency}\label{sec:dtclow}
\begin{figure}[!htb]
	\centering
	\includegraphics[width=1\columnwidth]{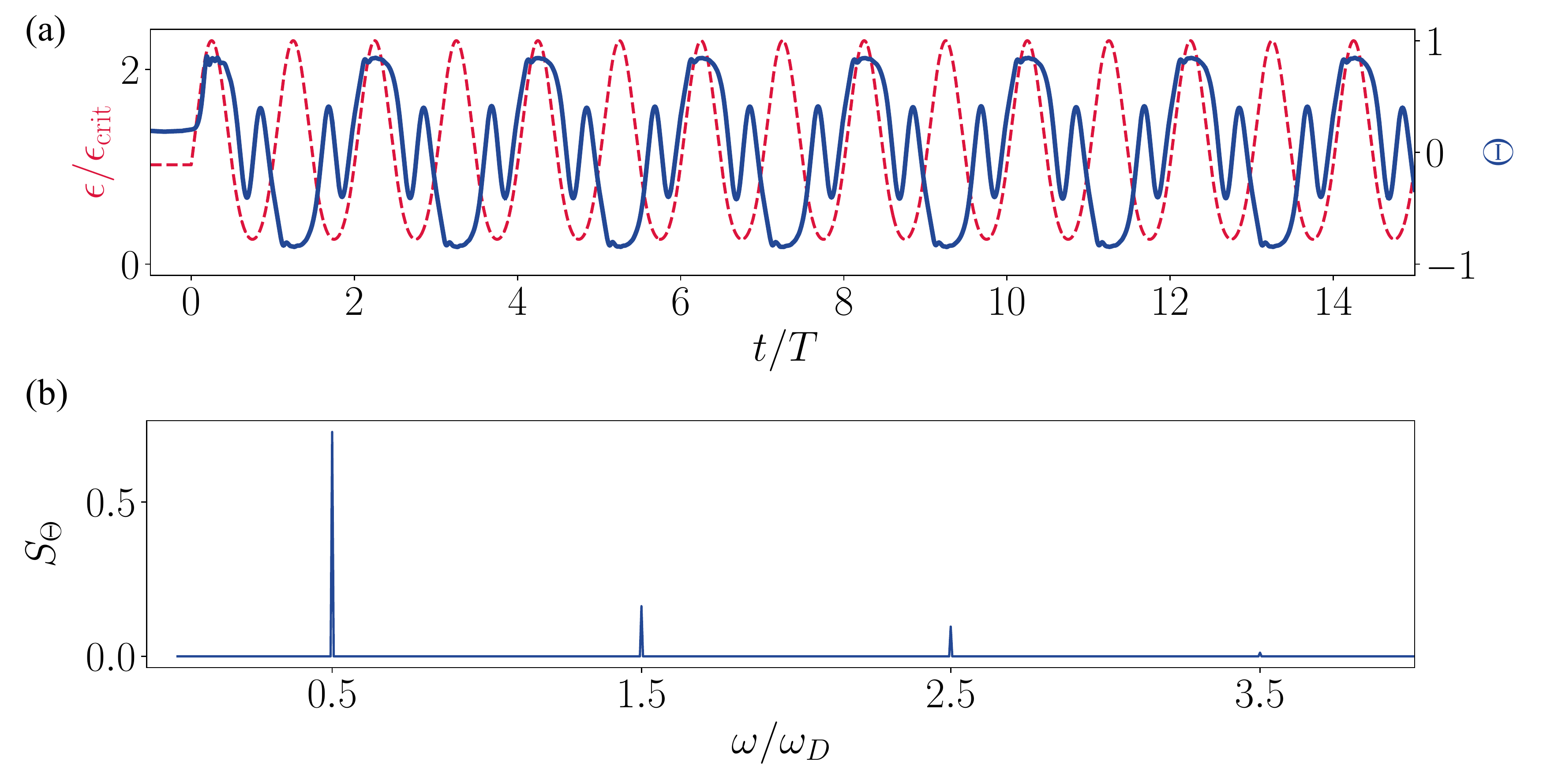}
	\caption{(a) Dependence of pump intensity and order parameter $\Theta$ on time for $\omega=0$ and $E_\mathrm{int}=0$. The driving parameters are $f_\mathrm{d}=0.5$ and $\omega_\mathrm{d}/2\pi =1$ kHz. (b) Corresponding power spectrum of $\Theta$ in (a).}
	\label{fig:crownDTC} 
\end{figure} 
In Fig.~\ref{fig:crownDTC}, we present an example of a DTC found in low driving frequencies and large driving strength. As briefly discussed in the main text, this unique DTC phase is marked by subdominant third harmonic oscillations of the period-doubling response. 

\section{DTC for varying driving strength}\label{sec:dtcfd}

\begin{figure}[!htb]
	\centering
	\includegraphics[width=1\columnwidth]{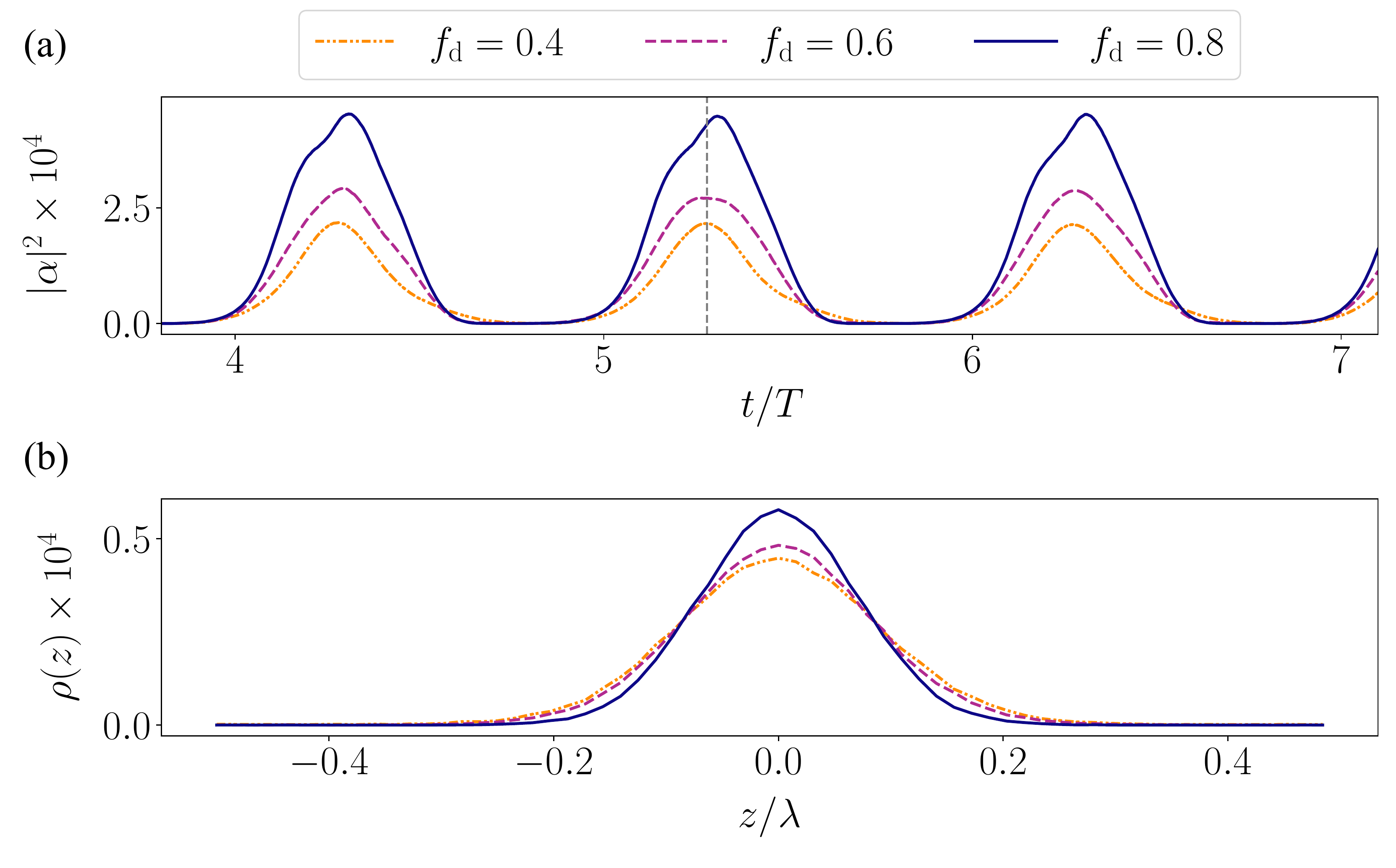}
	\caption{(a) Time evolution of the intracavity photon number for varying driving strength $f_\mathrm{d}$. (b) Snapshots of the single-particle density distribution of the atoms over a unit cell of the density wave. The snapshots are taken at the time denoted by the vertical dashed line in (a). The driving frequency is $\omega_\mathrm{d}/2\pi = 4~\mathrm{kHz}$.}
	\label{fig:dtcfd} 
\end{figure}

We show in Fig.~\ref{fig:dtcfd} the dependence of the DTC on the driving strength for an ideal atom-cavity system, where both short-range interaction and harmonic trap are neglected. The amplitude of oscillations in the intracavity field dynamics increases with $f_\mathrm{d}$ suggesting that the density modulations in the DW phase becomes more prominent, see Fig.~\ref{fig:dtcfd}(b).

\section{Coupling of momentum states due to the harmonic trap}\label{sec:coup}

An inhomogeneous single-particle potential leads to coupling of various single-particle momentum states. In the case of the harmonic trap considered here, this can be seen by going to the momentum space for the following single-particle Hamiltonian
\begin{align}
\hat{H}_{\mathrm{HO}} &= \int dz \hat{\Psi}^{\dagger}(z) V(z) \hat{\Psi}(z),
\end{align}
where the potential is
\begin{equation}
V(z) = \frac{m\omega^2_\mathrm{rec} b^2 z^2}{2}.
\end{equation}
The harmonic trap frequency relative to the recoil frequency is given by $b = \lambda^2/(2\pi^2 l^2_z)$, which follows from $\omega = \hbar/(l^2_zm)$ and $\omega_{\mathrm{rec}} = 2\pi^2\hbar/(m\lambda^2)$. We expand the field operators according to $\hat{\Psi}(z) = \sum_k \exp(i k z) \hat{a}_k$ and obtain
\begin{align}
\hat{H}_{\mathrm{HO}} &= \sum_{k_1,k_2} V(k_1- k_2) \hat{a}^{\dagger}_{k_1} \hat{a}_{k_2}, \label{eq:momH}
\end{align}
where
\begin{equation}\label{eq:potK}
V(k) = \int dz \exp(ikz) V(z).
\end{equation}
\begin{figure}[!htb]
	\centering
	\includegraphics[width=1\columnwidth]{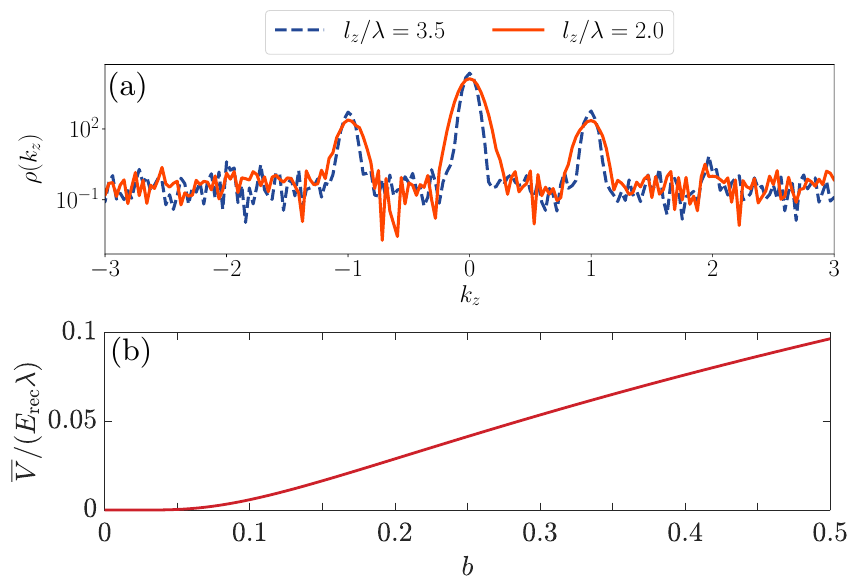}
	\caption{(a) Snapshots of the momentum distributions for a stable DTC ($f_\mathrm{d} = 0.3$) and a metastable DTC  ($f_\mathrm{d} = 0.6$). The remaining parameters $l_z/\lambda = 3.5$, $\omega_\mathrm{d}/2\pi = 3.5~\mathrm{kHz}$, and zero contact interaction, $E_{\mathrm{int}}=0$. (b) Effective coupling induced by the harmonic trap for momentum states with difference of $\Delta k = \pi/\lambda$ as a function of the unitless trap frequency $b$.}
	\label{fig:Veff} 
\end{figure} 

We can infer from Eq.~\eqref{eq:momH} that different pairs of momentum states can be coupled depending on the potential Eq.~\eqref{eq:potK}. To gain further analytical insight, we approximate the harmonic potential via
\begin{align}
V_\mathrm{eff}(z) &= \hbar\omega_\mathrm{rec} b \left(1- \exp\left(-\frac{z^2}{2 l^2_z}\right) \right) \\
&= V(z) + \mathcal{O}(z^4).
\end{align}
Note that this effective potential underestimates the effect of the actual harmonic potential which leads to an underestimation of the coupling strength between the momentum modes. Nevertheless, we will use this Gaussian potential to obtain an analytical expression for $V(k)$. Taking the Fourier transform of $V_\mathrm{eff}(z)$ yields
\begin{equation}\label{eq:Veffk}
V(\Delta k ) = \hbar \omega_\mathrm{rec} \lambda \frac{\sqrt{b}}{\pi\sqrt{2}}\exp(-(\Delta k)^2\, l^2_z/2),
\end{equation}
The presence of the harmonic trap creates additional momentum excitations and it broadens the distribution around the relevant ones, namely $|{k_0}\rangle\equiv|\mathbf{k}=0\rangle$ and $|{k_1}\rangle\equiv|\mathbf{k}=\pm 2\pi/\lambda\rangle$, as demonstrated in Fig.~\ref{fig:Veff}(a). The coupling between these two modes is crucial in the formation of both DW and DTC phases. If other momentum modes become significantly coupled to these two modes, then the DTC phase may become unstable as atoms occupy other momentum modes. To get an order of magnitude estimate for the strength of harmonic confinement that may lead to instability of the DTC phase, we calculate the effective coupling strength between the relevant momentum modes and a momentum state between them, i.e., $\Delta k = (k_1-k_0)/2 = \pi/\lambda$. Using Eq.~\eqref{eq:Veffk}, we get
\begin{equation}
\overline{V} \equiv V(\Delta k = \pi/\lambda) = \hbar \omega_\mathrm{rec} \lambda \frac{\sqrt{b}}{\pi\sqrt{2}}\exp\left(-1/4b\right).
\end{equation}

In Fig.~\ref{fig:Veff}(b), we show the dependence of the trap-induced coupling between $|k_0\rangle$ and $|k_1\rangle$ on $b$. The coupling is negligible for $b < 0.05$, which corresponds to a ratio between the pump wavelength and oscillator length of $\lambda/l_z < 1$. We then get an order of magnitude condition $l_z > \mathcal{O}(\lambda)$ for which the momentum modes $|k_0\rangle$ and $|k_1\rangle$ are mainly coupled by the cavity and the effect of the trap remains minimal. That is, we do not expect to observe any stable DTC for $l_z < \mathcal{O}(\lambda)$ when the trap-induced coupling becomes significant.
\bibliography{biblio}

\end{document}